\documentclass[camera,letterpaper,nomarginnotes,nonarrowgutter]{jpaper}
\usepackage{listings}
\usepackage{fancyhdr}
\usepackage{datetime}
\usepackage{cite}
\usepackage{amsmath,amssymb,amsfonts}
\usepackage{algorithmic}
\usepackage{graphicx}
\usepackage{textcomp}
\usepackage[table]{xcolor}
\usepackage{tikz}
\usepackage[utf8]{inputenc}
\usepackage[T1]{fontenc}
\usepackage{booktabs} %
\usepackage{setspace}
\usepackage[italic]{mathastext}
\usepackage{array}
\usepackage{titlesec}
\usepackage[normalem]{ulem}
\usepackage{multirow}
\usepackage{multicol}
\usepackage{color}
\usepackage[font={small,bf}]{caption}
\usepackage{float}
\usepackage[font={small,bf}]{subcaption}
\usepackage[linesnumbered,ruled]{algorithm2e}
\usepackage{makecell}
\usepackage{pifont}
\usepackage[]{microtype}

\usepackage[colorinlistoftodos,prependcaption,textsize=small]{todonotes}
\usepackage{marginnote} 

\usepackage{clipboard}
\usepackage{balance}
\usepackage[framemethod=tikz]{mdframed}
\usepackage{xargs}
\usepackage[most]{tcolorbox}

\usepackage[bookmarks=true,breaklinks=true,colorlinks,linkcolor=black,citecolor=blue,urlcolor=black]{hyperref}
\usepackage[resetlabels]{multibib}

\definecolor{bestresult}{HTML}{b3e2cd} %

\definecolor{denim}{rgb}{0.08, 0.38, 0.74}
\definecolor{darkolivegreen}{rgb}{0.33, 0.42, 0.18}
\definecolor{dgreen}{rgb}{0.00, 0.75, 0.00}
\definecolor{darkpink}{rgb}{0.88, 0.28, 0.54}
\definecolor{forestgreen}{rgb}{0.0, 0.27, 0.13}
\definecolor{amber}{rgb}{1.0, 0.49, 0.0}
\definecolor{lightyellow}{rgb}{0.980, 0.956, 0.623}
\definecolor{lightblue}{rgb}{0.980, 0.956, 0.623}
\definecolor{darkamber}{rgb}{0.5, 0.19, 0.0}
\definecolor{dkgreen}{rgb}{0,0.6,0}
\definecolor{gray}{rgb}{0.5,0.5,0.5}
\definecolor{mauve}{rgb}{0.58,0,0.82}
\definecolor{lightmauve}{rgb}{0.68,0.4,0.92}
\definecolor{chocolate}{rgb}{0.48, 0.25, 0.0}
\definecolor{dollarbill}{rgb}{0.52,0.73,0.4}
\definecolor{dkdkgreen}{rgb}{0,0.45,0}
\definecolor{gfored}{rgb}{0.580, 0.050, 0.211}
\definecolor{darkwarmgray}{rgb}{0.15, 0.050, 0.05}
\definecolor{ups-truck}{rgb}{0.53, 0.28, 0.21}

\makeatletter
\g@addto@macro{\normalsize}{%
  \setlength{\abovedisplayskip}{2pt plus 1pt minus 1pt}
  \setlength{\belowdisplayskip}{2pt plus 1pt minus 1pt}
  \setlength{\intextsep}{2pt plus 1pt minus 1pt}
  \setlength{\textfloatsep}{3pt plus 1pt minus 1pt}
  \setlength{\dbltextfloatsep}{3pt plus 1pt minus 1pt}
  \setlength{\skip\footins}{4pt plus 1pt minus 1pt}
}
\def\BibTeX{{\rm B\kern-.05em{\sc i\kern-.025em b}\kern-.08em
    T\kern-.1667em\lower.7ex\hbox{E}\kern-.125emX}}

\def\UrlBreaks{\do\/\do-\/\do.\/\do:}

\expandafter\def\expandafter\UrlBreaks\expandafter{\UrlBreaks
  \do\a\do\b\do\c\do\d\do\e\do\f\do\g\do\h\do\i\do\j
  \do\k\do\l\do\m\do\n\do\o\do\p\do\q\do\r\do\s\do\t
  \do\u\do\v\do\w\do\x\do\y\do\z\do\A\do\B\do\C\do\D
  \do\E\do\F\do\G\do\H\do\I\do\J\do\K\do\L\do\M\do\N
  \do\O\do\P\do\Q\do\R\do\S\do\T\do\U\do\V\do\W\do\X
  \do\Y\do\Z}
  
\newcommand{\squishlist}{
 \begin{list}{$\circ$}
  { \setlength{\itemsep}{0pt}
     \setlength{\parsep}{0pt}
     \setlength{\topsep}{0pt}
     \setlength{\partopsep}{0pt}
     \setlength{\leftmargin}{1em}
     \setlength{\labelwidth}{1em}
     \setlength{\labelsep}{0.5em} } }

\newcommand{\squishsublist}{
\begin{list}{$\rightarrow$}
 { \setlength{\itemsep}{0pt}
    \setlength{\parsep}{0pt}
    \setlength{\topsep}{-10em}
    \setlength{\partopsep}{-3pt}
    \setlength{\leftmargin}{1em}
    \setlength{\labelwidth}{1em}
    \setlength{\labelsep}{0.5em} } }

\newcommand{\squishend}{
  \end{list}  }

\usepackage{titlesec}
\titlespacing*{\section}{0pt}{0.3ex}{0.1ex} %
\titlespacing*{\subsection}{0pt}{0.3ex}{0.3ex} %
\titlespacing*{\subsubsection}{0pt}{0.3ex}{0.3ex} %

\newcommand\mech{RawHash2\xspace}

\newcommand{\rh}{{RawHash}\xspace}
\newcommand{\rhmin}{{RawHash2-Minimizer}\xspace}

\newcommand\unc{UNCALLED\xspace}
\newcommand\sig{Sigmap\xspace}

\newcommand\ltitle{RawHash2: Mapping Raw Nanopore Signals\\ Using Hash-Based Seeding and Adaptive Quantization\xspace}

\newcommand{\release}{\href{https://github.com/CMU-SAFARI/RawHash}{https://github.com/CMU-SAFARI/RawHash}\xspace}

\newcommand\avgthrU{$26.5\times$\xspace}

\newcommand\avgthrS{$19.2\times$\xspace}

\newcommand\avgthrR{$4.0\times$\xspace}
\newcommand\maxthrR{$9.9\times$\xspace}

\newcommand\avgAccRH{$10.57\%$\xspace}

\newcommand\maxAccRH{$20.25\%$\xspace}

\newcommand\avgseqR{$1.9\times$\xspace}

\newcommandx{\changev}[2][1=]{\todo[linecolor=blue,backgroundcolor=blue!25,bordercolor=blue,#1,size=\scriptsize]{#2}}

\let\oldmarginnote\marginnote

\renewcommand{\marginnote}[2][rectangle,draw,fill=blue!40,rounded corners]{%
        \oldmarginnote{%
        \tikz \node at (0,0) [#1]{#2};}%
        }
        
\newcommand{\boxbegin} {
	\begin{tcolorbox}[enhanced, frame hidden, colback=gray!50, breakable]
}

\newcommand{\boxend} {
	\end{tcolorbox}
}

\newcommand{\yboxbegin} {
	\begin{tcolorbox}[breakable, enhanced, frame hidden, colback=yellow!50]
}

\newcommand{\yboxend} {
	\end{tcolorbox}
}

\mdfdefinestyle{graybox}{
    splittopskip=0,%
    splitbottomskip=0,%
    frametitleaboveskip=0,
    frametitlebelowskip=0,
    skipabove=0,%
    skipbelow=0,%
    leftmargin=0,%
    rightmargin=0,%
    innertopmargin=0mm,%
    innerbottommargin=0mm,%
    roundcorner=1mm,%
    backgroundcolor=lightblue,
    hidealllines=true}
    
\mdfdefinestyle{graybox2}{
    splittopskip=0,%
    splitbottomskip=0,%
    frametitleaboveskip=0,
    frametitlebelowskip=0,
    skipabove=0,%
    skipbelow=0,%
    leftmargin=0,%
    rightmargin=0,%
    innertopmargin=2mm,%
    innerbottommargin=2mm,%
    roundcorner=2mm,%
    backgroundcolor=lightblue,
    hidealllines=true}

\newcommand{\bboxbegin}{
    \begin{mdframed}[style=graybox]
}

\newcommand{\bboxend}{
    \end{mdframed}
}

\newcommand{\yyboxbegin}{
    \begin{mdframed}[style=graybox2]
}

\newcommand{\yyboxend}{
    \end{mdframed}
}

\let\oldtableofcontents\tableofcontents%
\renewcommand\tableofcontents{
  \oldtableofcontents%
  \clearpage
}

\makeindex

\def\BibTeX{{\rm B\kern-.05em{\sc i\kern-.025em b}\kern-.08em
    T\kern-.1667em\lower.7ex\hbox{E}\kern-.125emX}}

\newcites{supp}{Supplementary References}

\begin{document}

\title{\ltitle}

\newcommand{\affilETH}[0]{\small {$$}}
\author{
\vspace{-18pt}\\%
{Can Firtina}\quad%
{Melina Soysal}\quad%
{Joël Lindegger}\quad%
{Onur Mutlu}\quad%
\vspace{-3pt}\\%
\affilETH\emph{ETH Z{\"u}rich}%
\vspace{-12pt}
}

\maketitle
\thispagestyle{plain}

\setstretch{0.83}

\begin{abstract}
\noindent \textbf{Summary:} Raw nanopore signals can be analyzed while they are being generated, a process known as real-time analysis. Real-time analysis of raw signals is essential to utilize the unique features that nanopore sequencing provides, enabling the early stopping of the sequencing of a read or the entire sequencing run based on the analysis. The state-of-the-art mechanism, \rh, offers the first hash-based efficient and accurate similarity identification between raw signals and a reference genome by quickly matching their hash values. In this work, we introduce \mech, which provides major improvements over \rh, including more sensitive quantization and chaining algorithms, weighted mapping decisions, frequency filters to reduce ambiguous seed hits, minimizers for hash-based sketching, and support for the R10.4 flow cell version and POD5 and SLOW5 file formats. Compared to \rh, \mech provides better F1 accuracy (on average by \avgAccRH and up to \maxAccRH) and better throughput (on average by \avgthrR and up to \maxthrR) than \rh. \\
\textbf{Availability and Implementation:} \mech is available at \release. We also provide the scripts to fully reproduce our results on our GitHub page. \\
\end{abstract}

\section{Introduction} \label{sec:introduction}

Nanopore technology can sequence long nucleic acid molecules up to more than two million bases at high throughput~\cite{jain_nanopore_2018}. As a molecule moves through a tiny pore, called a \emph{nanopore}, ionic current measurements are generated at a certain throughput (e.g., around 450 bases per second for DNA~\cite{kovaka_targeted_2021, zhang_real-time_2021}). These electrical measurements, known as \emph{raw signals}, can be used to 1)~identify individual bases in the molecule with computational techniques such as \emph{basecalling}~\cite{senol_cali_nanopore_2019} and 2)~analyze raw signals directly \emph{without} translating them to bases~\cite{firtina_rawhash_2023}.

\begin{figure}[tbh]
  \centering
  \includegraphics[width=0.85\columnwidth]{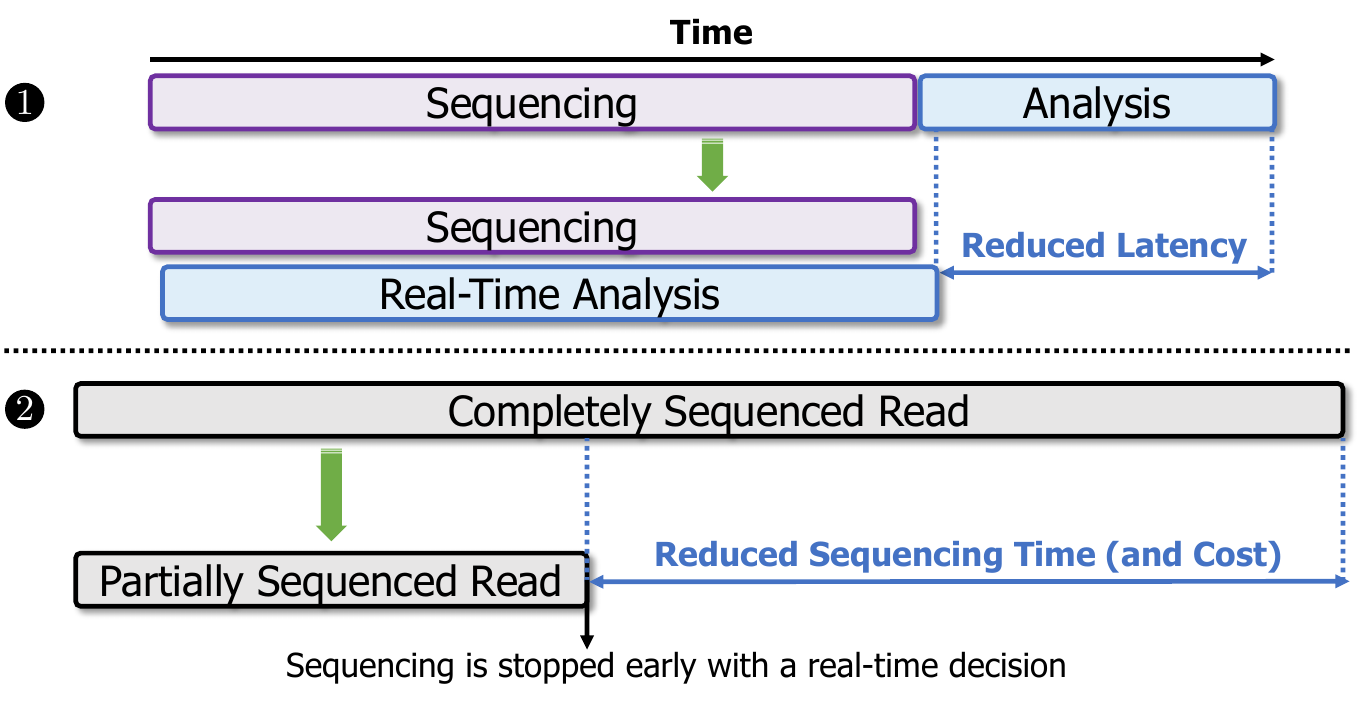}
  \caption{Two main benefits of real-time analysis with nanopore sequencing.}
  \label{fig:real-time}
\end{figure}

Computational techniques that can analyze the raw signals while they are generated at a speed that matches the throughput of nanopore sequencing are called \emph{real-time analysis}. Figure~\ref{fig:real-time} shows the two unique benefits that real-time analysis offers. First, real-time analysis allows for overlapping sequencing time with analysis time, as raw signals can be analyzed while they are being generated. Second, computational mechanisms can stop the sequencing of a read or the entire sequencing run early without sequencing the entire molecule or the sample using techniques known as Read Until~\cite{loose_real-time_2016} and Run Until~\cite{payne_readfish_2021}. The development of accurate and fast mechanisms for real-time analysis has the potential to significantly reduce the time and cost of genome analysis.

There are several mechanisms that can perform real-time analysis of raw nanopore signals to achieve accurate and fast genome analysis~\cite{edwards_real-time_2019, kovaka_targeted_2021, zhang_real-time_2021, payne_readfish_2021, dunn2021squigglefilter, bao_squigglenet_2021, shih_efficient_2023, sadasivan_rapid_2023, firtina_rawhash_2023, mikalsen_coriolis_2023, shivakumar_sigmoni_2023, lindegger_rawalign_2023}. Most of these solutions have three main limitations. First, many mechanisms offer limited scalability or support on resource-constrained devices due to their reliance on either 1)~deep neural networks (DNNs) for real-time base translation, which are usually computationally intensive and power-hungry~\cite{payne_readfish_2021, ulrich_readbouncer_2022}, or 2)~specialized hardware such as ASICs or FPGAs~\cite{bao_squigglenet_2021, dunn2021squigglefilter, shih_efficient_2023}. Second, while some mechanisms can directly analyze raw signals without base translation, offering an efficient alternative for real-time analysis~\cite{kovaka_targeted_2021, zhang_real-time_2021}, they often compromise accuracy or performance when applied to larger genomes. Third, methods based on machine learning often require retraining or reconfiguration~\cite{bao_squigglenet_2021, senanayake_deepselectnet_2023, sadasivan_rapid_2023}, adding a layer of complexity and reducing their flexibility for general use cases, such as read mapping to any genome.

Among the existing works, \rh~\cite{firtina_rawhash_2023} is the state-of-the-art mechanism that can accurately perform real-time mapping of raw nanopore signals for large genomes without translating them to bases with a hash-based seed-and-extend mechanism~\cite{altschul_basic_1990}. Despite its strengths in accuracy and performance, particularly for large genomes like the human genome, \rh exhibits several limitations that require further improvements.
First, \rh utilizes a simple quantization algorithm that assumes the raw signals are distributed uniformly across their normalized value range, which limits its efficiency and accuracy.
Second, \rh uses a chaining algorithm similar to that used in Sigmap~\cite{zhang_real-time_2021} without incorporating penalty scores used in minimap2~\cite{li_minimap2_2018}, which constrains its ability for more sensitive mapping.
Third, \rh performs chaining on all seed hits without filtering any of these seed hits, which substantially increases the workload of the chaining algorithm due to a large number of seed hits to chain.
Fourth, the decision-making mechanism in \rh for mapping reads to a reference genome in real-time relies on one of the mapping conditions being true (e.g., the ratio between the best and second-best chain scores), which makes it more prone to the outliers that can satisfy one of these conditions. A more robust and statistical approach that incorporates features beyond chaining scores can provide additional insights for making more sensitive and quick mapping decisions.
Fifth, while the hash-based mechanism in \rh is compatible with existing sketching techniques such as minimizers~\cite{roberts_reducing_2004, li_minimap2_2018}, strobemers~\cite{sahlin_effective_2021}, and fuzzy seed matching as in BLEND~\cite{firtina_blend_2023}, the benefits of these techniques are unknown for raw signal analysis as they are not used in \rh. Such evaluations could potentially provide additional insights on how to use the existing hash-based sketching techniques and reduce storage requirements while maintaining high accuracy.
Sixth, \rh lacks the support for recent advancements, such as the newer R10.4 flow cell version. The integration of these features can accelerate the adoption of both real-time and offline analysis.

In this work, our goal is to address the aforementioned limitations of \rh by improving its mechanism. To this end, we propose \mech to improve \rh in six directions.
First, to generate more accurate and unique hash values, we introduce a new quantization technique, \emph{adaptive quantization}. 
Second, to improve the accuracy of chaining and subsequently read mapping, we implement a more sophisticated chaining algorithm that incorporates penalty scores (as in minimap2). 
Third, to improve the performance of chaining by reducing its workload, \mech provides a filter that removes seeds frequently appearing in the reference genome, known as a \emph{frequency filter}.
Fourth, we introduce a statistical method that utilizes multiple features for making mapping decisions based on their weighted scores to eliminate the need for manual and fixed conditions to make decisions.
Fifth, we extend the hash-based mechanism to incorporate and evaluate the minimizer sketching technique, aiming to reduce storage requirements without significantly compromising accuracy.
Sixth, we integrate support for R10.4 flow cells and more recent file formats, POD5 and S/BLOW5~\cite{gamaarachchi_fast_2022}.

Compared to \rh, our extensive evaluations on five genomes of varying sizes and six different real datasets show that \mech provides higher accuracy (by \avgAccRH on average and \maxAccRH at maximum) and better read mapping throughput (by \avgthrR on average and \maxthrR at maximum). We make the following contributions:

\begin{itemize}
    \item We propose substantial algorithmic improvements to the state-of-the-art tool, \rh. These include 1)~more accurate quantization, 2)~more sensitive chaining with penalty scores, 3)~a frequency filter, 4)~mapping decisions based on a weighted sum of several features that can contribute to the decision, 5)~the minimizer sketching technique.
    \item We provide the support and evaluation for the newer flow cell version (i.e., R10.4) and file formats (i.e., POD5 and SLOW5).
\end{itemize}

\section{Methods} \label{sec:methods}

\rh is a mechanism to perform mapping between raw signals by quickly matching their hash values. We provide the details of the \rh mechanism in Supplementary Section~\ref{suppsec:rh_overview}. \mech provides substantial improvements over \rh in six key directions.
First, to generate more accurate and distinct hash values from raw signals, \mech improves the quantization mechanism with an \emph{adaptive} approach such that signal values are quantized non-uniformly based on the characteristics of a nanopore model.
Second, to provide more accurate mapping, \mech improves the chaining algorithm in \rh with more accurate penalty scores. 
Third, to reduce the workload in chaining for improved performance, we integrate a frequency filter to quickly eliminate the seed hits that occur too frequently. 
Fourth, to make more accurate and quick mapping decisions, \mech determines whether a read should be mapped at a specific point during sequencing by using a weighted sum of multiple features.
Fifth, to reduce the storage requirements of seeds, \mech incorporates and evaluates the benefits of minimizer sketching technique.
Sixth, \mech includes support for the latest features introduced by ONT, such as new file formats and flow cells.

\subsection{Adaptive Quantization} \label{subsec:adaptivequantization}

To improve the accuracy and uniqueness of hash values generated from raw nanopore signals, \mech introduces a new \emph{adaptive} quantization technique that we explain in four steps.

First, to enable a more balanced and accurate assignment of normalized signal values into quantized values (i.e., buckets), \mech performs a bifurcated approach to define two different ranges: 1)~fine range and 2)~coarse range. These ranges are useful for fine-tuning the boundaries of normalized signal values, $s$ falling into a certain quantized value, $q(s)$ within the integer value range $[0, n]$, as the normalized distribution of signal values is not uniform across all ranges.
Second, within the \emph{fine range}, normalized signal values are quantized into smaller intervals to enable a high resolution, $f_{r}$, for quantization due to the larger number of normalized signal values that can be observed within this range. The boundaries of the fine range, ($f_{min}$ and $f_{max}$), are empirically defined to enable robustness and high accuracy applicable given a flow cell (e.g., R9.4) and parameters to \mech to enable flexibility.
Third, the normalized signal values outside the fine range (i.e., the \emph{coarse range}) are quantized into larger intervals with low resolution, $c_{r} = (1 - f_{r}) \times 0.5$, to enable a more balanced load of quantized values across all ranges by assigning more signal values within this range into the same quantized value.
Fourth, depending on the range that a normalized signal is in, its corresponding quantized value is assigned as shown in Equation~\ref{eq:adaptivequant}. The adaptive quantization approach can enable a more balanced and accurate distribution of quantized values by better distinguishing closeby signal values with high resolution and grouping signals more efficiently in the coarser range.

\begin{equation}\label{eq:adaptivequant}
q(s) = \begin{cases} 
\lfloor n \times (f_{r} \times \frac{(s - f_{min})}{f_{max} - f_{min}}) & \text{if } f_{min} \leq s \leq f_{max} \\
\lfloor n \times (f_r + c_{r} \times s) & \text{if } s < f_{min} \\
\lfloor n \times (f_r + c_{r} + c_{r} \times s) & \text{if } s > f_{max} 
\end{cases}
\end{equation}

\subsection{Chaining with Penalty Scores}
To identify the similarities between a reference genome (i.e., target sequence) and a raw signal (i.e., query sequence), the series of seed hits within close proximity in terms of their matching positions are identified using a dynamic programming (DP) algorithm, known as \emph{chaining}. Using a chaining terminology similar to that of minimap2~\cite{li_minimap2_2018}, a seed hit between a reference genome and a raw signal is usually represented by a 3-tuple $(x,y,w)$ value, known as \emph{anchor}, where $w$ represents the length of the region that a seed spans, the start and end positions of a matching interval in a reference genome and a raw signal is represented by $[x-w+1,x]$ and $[y-w+1,y]$, respectively. The chain of anchors within close proximity is identified by calculating the optimal chain score $f(i)$ of each anchor $i$, where $f(i)$ is calculated based on predecessors of anchor $i$ when anchors are sorted by their reference positions. To calculate the chain score, $f(i)$, with dynamic programming, \rh performs the following computation as used in Sigmap~\cite{zhang_real-time_2021}.

\begin{equation}\label{eq:chainold}
f(i)=\max\big\{\max_{i>j\ge 1} \{ f(j)+\alpha(j,i)\},w_i\big\}
\end{equation}

where $\alpha(j,i)=\min\big\{\min\{y_i-y_j,x_i-x_j\},w_i\big\}$ is the length of the matching region between the two anchors. Although such a design is useful when identifying substantially fewer seed matches using a seeding technique based on distance calculation as used in Sigmap, \rh identifies a larger number of seed matches as it uses hash values to identify the matching region, which is usually faster than a distance calculation with the cost of reduced sensitivity.

To identify the correct mapping regions among such a large number of seed matches, \mech uses a more sensitive chaining technique as used in minimap2 by integrating the gap penalty scores such that the chain score of an anchor $i$ is calculated as shown in Equation~\ref{eq:chain}:

\begin{equation}\label{eq:chain}
f(i)=\max\big\{\max_{i>j\ge 1} \{ f(j)+\alpha(j,i)-\beta(j,i) \},w_i\big\}
\end{equation}

where $\beta(j,i)=\gamma_c\big((y_i-y_j)-(x_i-x_j)\big)$ is the penalty score calculated based on the gap distance, $l$, between a pair of anchors $i$ and $j$ where $\gamma_c(l) = 0.01\cdot w\cdot|l|+0.5\log_2|l|$. Based on the chain score calculation with gap costs, \mech integrates similar heuristics, mapping quality calculation, and the same complexity when calculating the chaining scores with the gap penalty as described in minimap2~\cite{li_minimap2_2018}.

\subsection{Frequency Filters}
\mech introduces a two-step frequency filtering mechanism to 1)~reduce the computational workload of the chaining process by limiting the number of anchors it processes and 2)~focus on more unique and potentially meaningful seed hits. First, to reduce the number of queries made to the hash table for identifying seed hits, \mech eliminates non-unique hash values generated from raw signals that appear more frequently than a specified threshold. Second, \mech evaluates the frequency of each seed hit within the reference genome and removes those that surpass a predefined frequency threshold, which reduces the overall workload of the chaining algorithm by providing a reduced set of more unique seed hits.

\subsection{Weighted Mapping Decision} \label{subsec:weightedmapping}

\rh performs mapping while receiving chunks of signals in real-time, as provided by nanopore sequencers. It is essential to decide if a read maps to a reference genome as quickly as possible to avoid unnecessary sequencing. The decision-making process in \rh is based on a series of conditional checks involving chain scores. These checks are performed in a certain order and against fixed ratios and mean values, making the decision mainly rigid and less adaptive to variations.

To employ a more statistical approach that can generalize various variations between different data sets and genomes, \mech calculates a weighted sum of multiple features that can impact the mapping decision. To achieve this, \mech calculates normalized ratios of several metrics based on mapping quality and chain scores. These metrics are 1)~the ratio of the mapping quality to a sufficiently high mapping quality (i.e., 30), 2)~mapping quality ratio between the best chain and the mean quality of all chains, and 3)~the ratio of the chain score between the best and the mean score of all chains. These ratios are combined into a weighted sum as follows: $w_{\text{sum}} = \sum_{i=1} r_i \times w_i$, where $r_{i}$ is a ratio of a particular metric, and $w_{i}$ is the weight assigned for that particular metric. The weighted sum, $w_{sum}$, is compared against a predefined threshold value to decide if a read is considered to be mapped. \mech maps a read if the weighted sum exceeds the threshold. Such a weighted sum approach allows \mech to adaptively consider multiple aspects of the data and eliminates the potential effect of the ordering of these checks to achieve improved mapping accuracy while maintaining computational efficiency.

\subsection{Minimizer Sketching} \label{subsec:minimizer}
\rh provides the opportunity to integrate the existing hash-based sketching techniques such as minimizers~\cite{roberts_reducing_2004, li_minimap2_2018} for 1)~reduced storage requirements of index in disk and memory and 2)~faster mapping due to fewer seed queries and hits.

To reduce the storage requirements of storing seeds in raw signals and due to their widespread application, \mech integrates minimizers in two steps. First, \mech generates hash values for seeds in both the reference genome and the raw signal. Second, within each window comprising $w$ hash values, the minimum hash value is selected as the minimizer. These minimizer hash values can be used to find similarities using hash tables (similar to \rh that uses hash values of all k-mers) while significantly reducing the number of hash values that need to be stored and queried during the mapping process as opposed to storing all k-mers.

\subsection{Support for New Data Formats and Flow Cells}\label{subsec:fileandr10methods}
To enable better and faster adoption, \mech incorporates support for 1)~recent data formats for storing raw signals, namely POD5 and SLOW5~\cite{gamaarachchi_fast_2022} as well as the existing FAST5 format, and 2)~the latest flow cell versions due to two main reasons. First, transitioning from the FAST5 to the POD5 file format is crucial for broad adoption, as POD5 is the new standard file format introduced by Oxford Nanopore Technologies (ONT). Second, integrating the newer flow cell versions is challenging as it requires optimization of parameters involved in mapping decisions as well as segmentation. \mech enables mapping the raw signals from R10.4 flow cells by optimizing the segmentation parameters for R10.4 and adjusting the scoring parameters involved in chaining settings to enable accurate mapping for R10.4 flow cells.

\section{Results} \label{sec:results}
\subsection{Evaluation Methodology} \label{subsec:evaluation}

We implement the improvements we propose in \mech directly on the \rh implementation. Similar to \rh, \mech provides the mapping information using a standard pairwise mapping format (PAF).

We compare \mech with the state-of-the-art works \unc~\cite{kovaka_targeted_2021}, \sig~\cite{zhang_real-time_2021}, \rh~\cite{firtina_rawhash_2023} in terms of throughput, accuracy, and the number of bases that need to be processed before stopping the sequencing of a read to estimate the benefits in sequencing time and cost. We provide the release versions of these tools in Supplementary Table~\ref{tab:versions}. For throughput, we calculate the number of bases that each tool can process per second per CPU thread, which is essential to determine if a calculation in a single thread is at least as fast as the speed of sequencing from a single nanopore (i.e., single pore). In many commonly used nanopore sequencers, a nucleic acid molecule passes through a pore at around 450 bases and 400 per second with sampling rates of 4 KHz and 5 KHz for DNA in R9.4.1 and R10.4.1, respectively~\cite{kovaka_targeted_2021, sam_kovaka_uncalled4_2024}. Since each read is mapped using a single thread for all tools, the throughput calculation is not affected by the number of threads available to these tools. Rather, this throughput calculation shows how many pores a single thread can process and how many CPU threads are needed to process the entire flow cell with many pores (e.g., 512 pores in a MinION flow cell). To show these results, we calculate 1)~the number of pores that a single thread can process by dividing throughput by the number of bases sequenced per second per single pore and 2)~the number of threads needed to cover the entire flow cell.

For accuracy, we analyze three use cases: 1)~read mapping, 2)~contamination analysis, and 3)~relative abundance estimation. To identify the correct mappings, we generate the ground truth mapping output in PAF by mapping the basecalled sequences of corresponding raw signals to their reference genomes using minimap2~\cite{li_minimap2_2018}. We use \texttt{UNCALLED pafstats} to compare the mapping output from each tool with their corresponding ground truth mapping output to calculate precision ($P = TP/(TP+FP)$), recall ($R = TP/(TP + FN)$), and F1 ($F1 = 2 \times (P \times R)/(P+R)$) values, similar to \rh~\cite{firtina_rawhash_2023}. For read mapping, we compare the tools in terms of their precision, recall, and F-1 scores. For contamination analysis, the goal is to identify if a particular sample is contaminated with a certain genome (or set of genomes), which makes the precision metric more important for such a use case. For this use case, we compare the tools in terms of their precision in the main paper and show the full results (i.e., precision, recall, and F1) in the Supplementary Table~\ref{supptab:accuracyfull}.
For relative abundance estimation, we calculate the abundance ratio of each genome based on the ratio of reads mapped to a particular genome compared to all read mappings. We calculate the Euclidean distance of each estimation to the ground truth estimations generated based on minimap2 mappings of corresponding basecalled reads. We estimate the relative abundances based on the number of mapped reads rather than the number of mapped bases as we identify that larger genomes usually require sequencing a larger number of bases to map a read, which can lead to skewed estimations towards larger genomes.

To estimate the benefits in sequencing time and the cost per read, we identify the average sequencing length before making the mapping decision for a read. For all of our analyses, we use the default parameters of each tool as we show in Supplementary Table~\ref{tab:parameters}. Supplementary Table~\ref{supptab:dataset} shows the real dataset details we use in our evaluation, including more details about sequencing run settings and flow cell versions (i.e., R9.4.1 and R10.4). For all the datasets except D7, we use already basecalled reads available with the raw electrical signals. For the D7 dataset, we basecall the raw signals using the Dorado basecaller. Although \mech does not use the minimizer sketching technique by default to achieve the maximum accuracy, we evaluate the benefits of minimizers in \mech, which we refer to as \texttt{\rhmin}. Since the evaluated versions of \unc, \sig, and \rh do not provide the support for R10.4 dataset, we show the corresponding results with the R10.4 dataset without comparing to these tools. When comparing \mech to other tools we always use FAST5 files containing raw signals from R9.4 flow cells on an isolated machine and SSD. We use AMD EPYC 7742 processor at 2.26GHz to run the tools. We use 32 threads for all the tools.

\subsection{Throughput} \label{subsec:perfmemory}

\begin{figure}[tbh]
  \centering
  \includegraphics[width=\columnwidth]{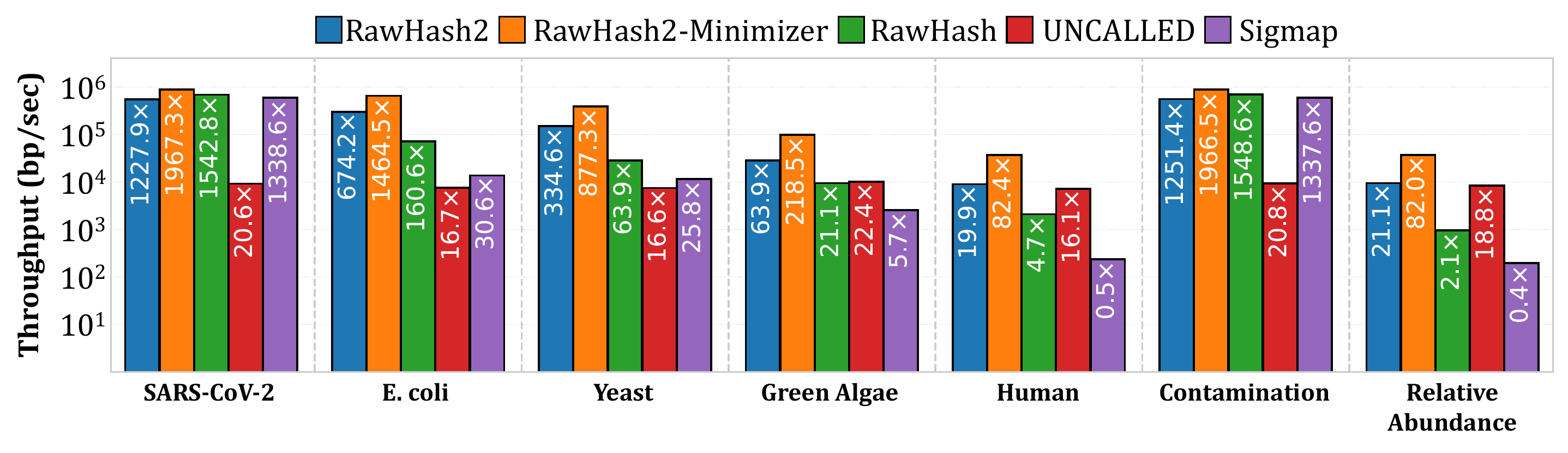}
  \caption{Throughput of each tool. Values inside the bars show how many nanopores (i.e., pores) that a single CPU thread can process.}
  \label{fig:throughput}
\end{figure}

Figure~\ref{fig:throughput} shows the results for 1)~throughput per single CPU thread and 2)~number of pores that a single CPU thread can analyze as annotated by the values inside the bars. We make three key observations.
First, we find that \mech provides average throughput \avgthrU, \avgthrS, and \avgthrR better than \unc, \sig, and \rh, respectively. Such a speedup, specifically over the earlier work \rh, is achieved by reducing the workload of chaining with the unique and accurate hash values using the new quantization mechanism and the filtering technique (see the filtering ratios in Supplementary Table~\ref{supptab:filtered_seed_hits}).
Second, we find that \rhmin enables reducing the computational requirements for mapping raw signals and enables improving the average throughput by $2.5\times$ compared to \mech, while the other computational resources, such as the peak memory usage and CPU time in both indexing and mapping, and the mean time spent per read are also significantly reduced as shown in Supplementary Tables~\ref{supptab:performance} and Supplementary Figure~\ref{suppfig:timeperread}.
Third, \rhmin requires \emph{at most} 7 threads for analyzing the entire flowcell for any evaluated dataset, while \mech requires at most 2 threads for smaller genomes and 9 to 26 threads for Green Algae and human. This shows that \mech and \rhmin can reduce computational requirements and energy consumption significantly compared to 28 threads required, on average, regardless of the genome size for \unc, which is critical for portable sequencing.
We conclude that \mech and \rhmin significantly reduce the computational overhead of mapping raw signals to reference genomes, enabling better scalability to even larger genomes.

\subsection{Accuracy} \label{subsec:accuracy}

Table~\ref{tab:accuracy} shows the accuracy results for read mapping, contamination analysis, and relative abundance estimation based on their corresponding most relevant accuracy metrics (results with all metrics are shown in Supplementary Table~\ref{supptab:accuracyfull} and Supplementary Figure~\ref{suppfig:accuracy}). We make two key observations. First, we find that \mech provides 1)~the best accuracy in terms of the F1 score in all datasets for read mapping, 2)~the best precision for contamination analysis, and 3)~the most accurate relative abundance estimation. This is mainly achieved because 1)~the adaptive quantization enables finding more accurate mapping positions while substantially reducing the false seed hits due to less precise quantization in \rh, and 2)~the more sensitive chaining implementation with penalty scores can identify the correct mappings more accurately.
Second, \rhmin provides mapping accuracy similar to that of \mech with an exception for the human genome and better accuracy than \rh, providing substantially better performance results as discussed in Section~\ref{subsec:perfmemory}. Such an accuracy-performance trade-off puts \rhmin in an important position when a slight drop in accuracy can be tolerated for a particular use case when a substantially better throughput is needed.
For the relatively lower accuracy that \mech and \rhmin achieve compared to minimap2, we believe the accuracy gap is due to the increased difficulty in distinguishing the chain with the correct mapping position among many chains with similar quality scores, potentially due to the false seed matches in repetitive regions. Although our in-house evaluation shows that accuracy can substantially be improved further by enabling the correct chains to be distinguished more accurately than the incorrect chains with more sensitive quantization parameters, this comes with increased performance costs due to increased seed matches and chaining calculations. Future work can focus on designing more sensitive filters to improve the accuracy for larger and repetitive genomes by eliminating seed matches from such false regions. We conclude that \mech is the most accurate tool regardless of the genome size, while the minimizer sketching technique in \rhmin can provide better accuracy than \rh and on-par accuracy to all other tools while providing the best overall performance.

\begin{table}[tbh]
\centering
\caption{Accuracy.}
\resizebox{\columnwidth}{!}{\begin{tabular}{@{}llrrrrr@{}}\toprule
\textbf{Dataset} & \textbf{Metric} & \textbf{RH2} & \textbf{RH2-Min.} & \textbf{RH} & \textbf{UNCALLED} & \textbf{Sigmap} \\\midrule
SARS-CoV-2 & F1 & \cellcolor{bestresult}\textbf{0.9867} & 0.9691 & 0.9252 & 0.9725 & 0.7112 \\
E. coli & F1 & \cellcolor{bestresult}\textbf{0.9748} & 0.9631 & 0.9280 & 0.9731 & 0.9670 \\
Yeast & F1 & \cellcolor{bestresult}\textbf{0.9602} & 0.9472 & 0.9060 & 0.9407 & 0.9469 \\
Green Algae & F1 & \cellcolor{bestresult}\textbf{0.9351} & 0.9191 & 0.8114 & 0.8277 & 0.9350 \\
Human & F1 & \cellcolor{bestresult}\textbf{0.7599} & 0.6699 & 0.5574 & 0.3197 & 0.3269 \\
\midrule
Contamination & Precision & \cellcolor{bestresult}\textbf{0.9595} & 0.9424 & 0.8702 & 0.9378 & 0.7856 \\
\midrule
Rel. Abundance & Distance & \cellcolor{bestresult}\textbf{0.2678} & 0.4243 & 0.4385 & 0.6812 & 0.5430 \\
\bottomrule
\multicolumn{7}{l}{\footnotesize Best results are \colorbox{bestresult}{\textbf{highlighted}}.} \
\end{tabular}
}
\label{tab:accuracy}
\end{table}

\subsection{Sequencing Time and Cost} \label{subsec:sequencingcost}

Table~\ref{tab:seq_bases} shows the average sequencing lengths in terms of bases and chunks that each tool needs to process before stopping the sequencing process of a read. Processing fewer bases can significantly help reduce the overall sequencing time and potentially the cost spent for each read by enabling better utilization of nanopores without sequencing the reads unnecessarily. We make three key observations.
First, \mech reduces the average sequencing length by \avgseqR compared to \rh mainly due to the improvements in mapping accuracy, which enables making quick decisions without using longer sequences.
Second, as the genome size increases, \mech provides the smallest average sequencing lengths compared to all tools.
Third, when the average length of sequencing is combined with other important metrics such as mapping accuracy in terms of F1 score and throughput, \mech provides the best trade-off in terms of all these three metrics for all datasets as shown in Supplementary Figure~\ref{suppfig:combined}. We conclude that \mech is the best tool for longer genomes to reduce the sequencing time and cost per read as it provides the smallest average sequencing lengths, while \unc is the best tool for shorter genomes.

\begin{table}[tbh]
\caption{Average length of sequencing per read.}
\resizebox{\columnwidth}{!}{\begin{tabular}{@{}lrrrrr@{}}\toprule
\textbf{Dataset} & \textbf{RH2} & \textbf{RH2-Min.} & \textbf{RH} & \textbf{UNCALLED} & \textbf{Sigmap} \\\midrule
SARS-CoV-2 & 443.92  & 460.85  & 513.95  & \cellcolor{bestresult}\textbf{184.51}  & 452.38 \\
E. coli & 851.31  & 1,030.74  & 1,376.14  & \cellcolor{bestresult}\textbf{580.52}  & 950.03 \\
Yeast & \cellcolor{bestresult}\textbf{1,147.66}  & 1,395.87  & 2,565.09  & 1,233.20  & 1,862.69 \\
Green Algae & \cellcolor{bestresult}\textbf{1,385.59}  & 1,713.46  & 4,760.59  & 5,300.15  & 2,591.16 \\
Human & \cellcolor{bestresult}\textbf{2,130.59}  & 2,455.99  & 4,773.58  & 6,060.23  & 4,680.50 \\
\midrule
Contamination & 670.69  & \cellcolor{bestresult}\textbf{667.89}  & 742.56  & 1,582.63  & 927.82 \\
\midrule
Rel. Abundance & \cellcolor{bestresult}\textbf{1,024.28}  & 1,182.04  & 1,669.46  & 2,158.50  & 1,533.04 \\
\bottomrule
\multicolumn{6}{l}{\footnotesize Best results are \colorbox{bestresult}{\textbf{highlighted}}.} \
\end{tabular}
}
\label{tab:seq_bases}
\end{table}

\subsection{Evaluating New File Formats and R10.4}\label{subsec:r104}
In Supplementary Table~\ref{supptab:pod5_fast5_resources} and \ref{supptab:r10accuracy}, we show the results when using different file formats for storing raw signals (i.e., FAST5, POD5, and BLOW5) and R10.4, respectively. We make two key observations.
First, we find that POD5 and SLOW5 significantly speed up total elapsed time compared to FAST5. These results indicate that a large portion of the overhead spent for reading from a file can be mitigated with approaches that can perform faster compression and decompression, as these signal files are mostly stored in a compressed form.
Second, we find that \mech can perform fast analysis with reasonable accuracy that can be useful for certain use cases (e.g., contamination analysis) when using raw signals from R10.4, although \mech achieves lower accuracy with R10.4 than using R9.4. This is likely because 1)~we use a k-mer model optimized for the R10.4.1 flow cell version rather than R10.4, and 2)~minimap2 can provide more accurate mapping due to improved accuracy of these basecalled reads. Future work can focus on generating a k-mer model specifically designed for R10.4 to generate more accurate results. We exclude the accuracy results for R10.4.1 as the number of events found for R10.4.1 is around $35\%$ larger than that of R10.4, which leads to inaccurate mapping. We suspect that our segmentation algorithm and parameters are not optimized for R10.4.1. Our future work will focus on improving these segmentation parameters and techniques to achieve higher accuracy with R10.4.1 as well as RNA sequencing data. We believe this can be achieved because \mech 1)~is highly flexible to change all the parameters corresponding to segmentation and 2)~can map accurately without requiring long sequencing lengths (Table~\ref{tab:seq_bases}), which can mainly be useful for RNA read sets.
We conclude that \mech can provide accurate and fast analysis when using the recent features released by ONT.

\section{Conclusion} \label{sec:conclusion}

We introduce \mech, a tool that provides substantial improvements over the previous state-of-the-art mechanism \rh. We make five key improvements over \rh: 1)~more sensitive quantization and chaining, 2)~reduced seed hits with filtering mechanisms, 3)~more accurate mapping decisions with weighted decisions, 4)~the first minimizer sketching technique for raw signals, and 5)~integration of the recent features from ONT.
We find the \mech provides substantial improvements in throughput and accuracy over \rh. We conclude that \mech, overall, is the best tool for mapping raw signals due to its combined benefits in throughput, accuracy, and reduced sequencing time and cost per read compared to the existing mechanisms, especially for longer genomes.

\section*{Acknowledgments}

We thank all members of the SAFARI Research Group for the stimulating and scholarly intellectual environment they provide. We acknowledge the generous gift funding provided by our industrial partners (especially by Google, Huawei, Intel, Microsoft, VMware), which has been instrumental in enabling the decade+ long research we have been conducting on accelerating genome analysis. This work is also partially supported by the Semiconductor Research Corporation (SRC), the European Union’s Horizon programme for research and innovation [101047160 - BioPIM] and the Swiss National Science Foundation (SNSF) [200021\_213084].

\setstretch{0.75}
\bibliographystyle{IEEEtran}
\bibliography{main}

\setstretch{1}
\onecolumn
\setcounter{secnumdepth}{3}
\clearpage
\begin{center}
\textbf{\LARGE Supplementary Material for\\ \ltitle}
\end{center}
\setcounter{section}{0}
\setcounter{equation}{0}
\setcounter{figure}{0}
\setcounter{table}{0}
\setcounter{page}{1}
\makeatletter
\renewcommand{\theequation}{S\arabic{equation}}
\renewcommand{\thetable}{S\arabic{table}}
\renewcommand{\thefigure}{S\arabic{figure}}
\renewcommand{\thesection}{\Alph{section}}
\renewcommand{\thesubsection}{\thesection.\arabic{subsection}}
\renewcommand{\thesubsubsection}{\thesubsection.\arabic{subsubsection}}

\setstretch{0.95}
\newcommand{\TextUnderscore}{\rule{.4em}{.4pt}}
\section{\rh Overview}\label{suppsec:rh_overview}

\mech builds improvements over \rh~\citesupp{supp_firtina_rawhash_2023}, a mechanism that provides the first hash-based similarity identification between a raw signal and a reference genome accurately and quickly. We show the overview of \rh in Supplementary Figure~\ref{suppfig:overview}. \rh has four key steps. First, to generate sequences of signals that can be compared to each other, \rh generates signals of k-mers, called \emph{events}, from both a reference genome and raw signals. To generate events from reference genomes, it uses a lookup table, called \emph{k-mer model}, that provides the expected signal value (i.e., event value) as a floating value for each possible k-mer where k is usually 6 or 9, depending on the flow cell version. To identify events (i.e., k-mers) in raw signals, \rh performs a segmentation technique to detect the abrupt changes in signals, which enables identifying the regions in signals generated when sequencing a particular k-mer. \rh uses the average value of signals within the same region as an event value. Due to the variations and noise in nanopore sequencing, event values can slightly differ from each other although they correspond to the same k-mer, making it challenging to directly match the event values to each other to identify matching k-mers between a reference genome and raw signals.

Second, to mitigate this noise issue, \rh quantizes the event values such that slightly different event values can be quantized into the same value to enable direct matching of quantized event values between a reference genome and raw signals.

Third, to reduce the number of potential matches without reducing accuracy, \rh concatenates the quantized event values of consecutive events (i.e., consecutive k-mers) and generates a hash value from these concatenated values.

Fourth, for the reference genome, these hash values are stored in a hash table along with their position information, which is usually known as the indexing step in read mapping. \rh uses the hash values of raw signals to query the previously constructed hash table to identify matching hash values, known as \emph{seed hits}, between a reference genome and a raw signal, which is then followed by chaining and mapping based on the seed hits.

\begin{figure}[tbh]
  \centering
  \includegraphics[width=0.7\columnwidth]{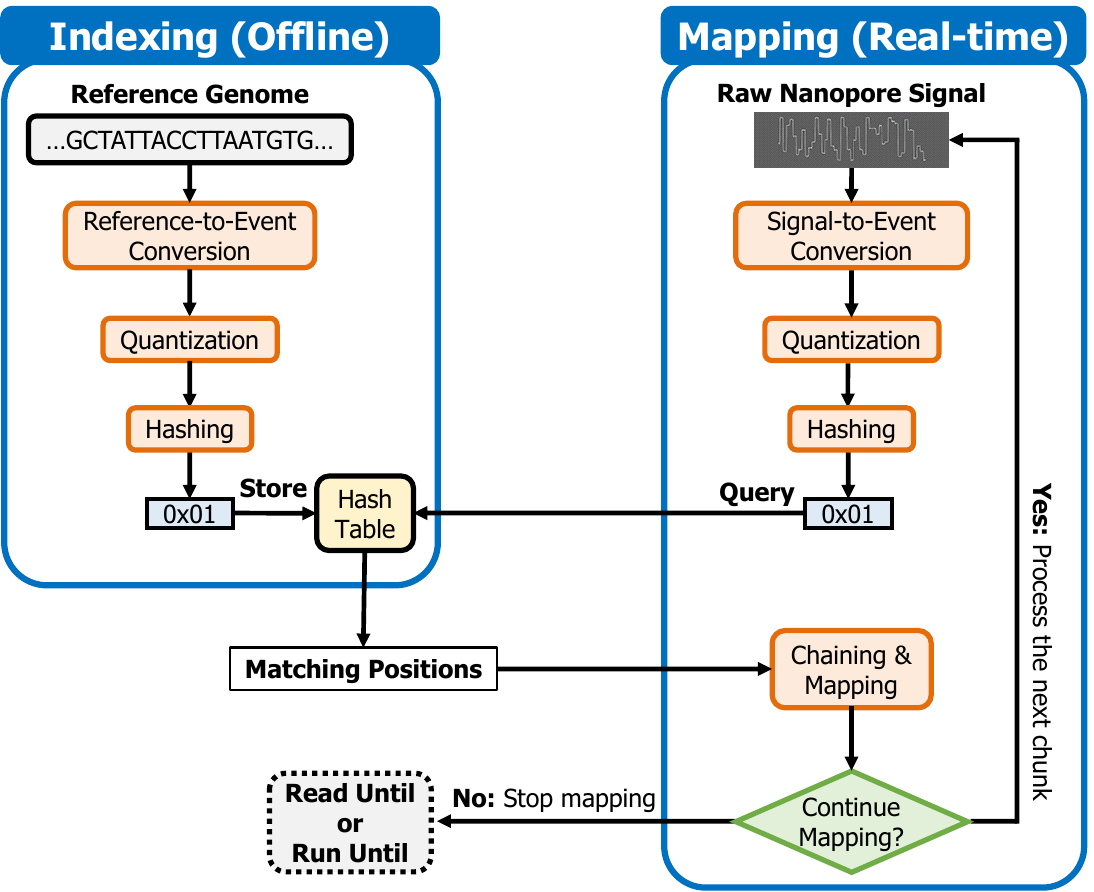}
  \caption{Overview of \rh.}
  \label{suppfig:overview}
\end{figure}

\clearpage
\section{Accuracy} \label{suppsec:accuracy}
\subsection{Read Mapping Accuracy} \label{suppsubsec:accuracyfull}

In Supplementary Table~\ref{supptab:accuracyfull}, we show the read mapping accuracy in all metrics (i.e., F1, Precision, and Recall) for all datasets. In Figure~\ref{suppfig:accuracy}, we show the same results as reported in Supplementary Table~\ref{supptab:accuracyfull} for visualizing the comparisons between tools and the trade-offs between each accuracy metric in all datasets.

\begin{table}[tbh]
\centering
\caption{Read mapping accuracy in all metrics: F1, Precision, and Recall.}
\begin{tabular}{@{}llrrrrr@{}}\toprule
\textbf{Dataset} & \textbf{Metric} & \textbf{RH2} & \textbf{RH2-Min.} & \textbf{RH} & \textbf{UNCALLED} & \textbf{Sigmap} \\\midrule
\multirow{3}{*}{SARS-CoV-2} & F1 & \cellcolor{bestresult}\textbf{0.9867} & 0.9691 & 0.9252 & 0.9725 & 0.7112 \\
 & Precision & \cellcolor{bestresult}\textbf{0.9939} & 0.9868 & 0.9832 & 0.9547 & 0.9929 \\
 & Recall & 0.9796 & 0.9521 & 0.8736 & \cellcolor{bestresult}\textbf{0.9910} & 0.5540 \\
\midrule
\multirow{3}{*}{E. coli} & F1 & \cellcolor{bestresult}\textbf{0.9748} & 0.9631 & 0.9280 & 0.9731 & 0.9670 \\
 & Precision & \cellcolor{bestresult}\textbf{0.9904} & 0.9865 & 0.9563 & 0.9817 & 0.9842 \\
 & Recall & 0.9597 & 0.9408 & 0.9014 & \cellcolor{bestresult}\textbf{0.9647} & 0.9504 \\
\midrule
\multirow{3}{*}{Yeast} & F1 & \cellcolor{bestresult}\textbf{0.9602} & 0.9472 & 0.9060 & 0.9407 & 0.9469 \\
 & Precision & 0.9553 & 0.9561 & 0.9852 & 0.9442 & \cellcolor{bestresult}\textbf{0.9857} \\
 & Recall & \cellcolor{bestresult}\textbf{0.9652} & 0.9385 & 0.8387 & 0.9372 & 0.9111 \\
\midrule
\multirow{3}{*}{Green Algae} & F1 & \cellcolor{bestresult}\textbf{0.9351} & 0.9191 & 0.8114 & 0.8277 & 0.9350 \\
 & Precision & 0.9284 & 0.9280 & 0.9652 & 0.8843 & \cellcolor{bestresult}\textbf{0.9743} \\
 & Recall & \cellcolor{bestresult}\textbf{0.9418} & 0.9104 & 0.6999 & 0.7779 & 0.8987 \\
\midrule
\multirow{3}{*}{Human} & F1 & \cellcolor{bestresult}\textbf{0.7599} & 0.6699 & 0.5574 & 0.3197 & 0.3269 \\
 & Precision & 0.8675 & 0.8511 & \cellcolor{bestresult}\textbf{0.8943} & 0.4868 & 0.4288 \\
 & Recall & \cellcolor{bestresult}\textbf{0.6760} & 0.5523 & 0.4049 & 0.2380 & 0.2642 \\
\midrule
\multirow{3}{*}{Contamination} & F1 & 0.9614 & 0.9317 & 0.8718 & \cellcolor{bestresult}\textbf{0.9637} & 0.6498 \\
 & Precision & \cellcolor{bestresult}\textbf{0.9595} & 0.9424 & 0.8702 & 0.9378 & 0.7856 \\
 & Recall & 0.9632 & 0.9212 & 0.8736 & \cellcolor{bestresult}\textbf{0.9910} & 0.5540 \\
\midrule
\multirow{3}{*}{Rel. Abundance} & F1 & \cellcolor{bestresult}\textbf{0.4659} & 0.3375 & 0.3045 & 0.1249 & 0.2443 \\
 & Precision & \cellcolor{bestresult}\textbf{0.4623} & 0.3347 & 0.3018 & 0.1226 & 0.2366 \\
 & Recall & \cellcolor{bestresult}\textbf{0.4695} & 0.3404 & 0.3071 & 0.1273 & 0.2525 \\
\bottomrule
\multicolumn{7}{l}{\footnotesize Best results are \colorbox{bestresult}{\textbf{highlighted}}.} \\
\end{tabular}

\label{supptab:accuracyfull}
\end{table}

\begin{figure}[tbh]
\centering
\includegraphics[width=0.8\columnwidth]{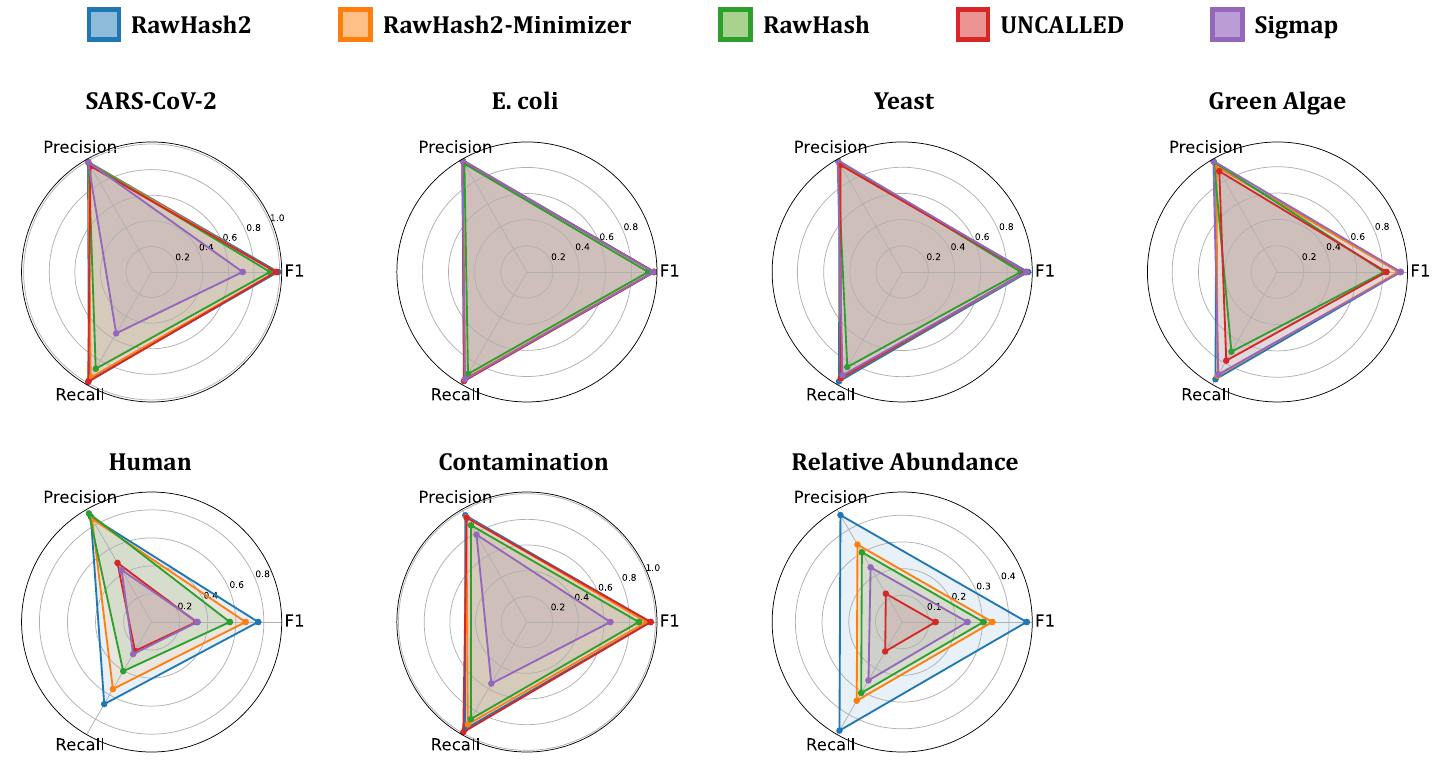}
\caption{Read mapping accuracy results in terms of F1 score, precision, and recall across different datasets. The dotted triangles show the best possible results, where each edge shows the best result for its corresponding metric.}
\label{suppfig:accuracy}
\end{figure}

\clearpage

\subsection{R10.4 Accuracy and Performance} \label{suppsubsec:r10accuracy}

In Supplementary Table~\ref{supptab:r10accuracy}, we show the accuracy and performance results in terms of throughput and mean time spent per read when using R10.4 flow cells. For comparison purposes between R10.4 and R9.4, we include the results from R9.4 flow cells for \emph{E. coli}. We do not show the R9.4 results for \emph{S. aureus}, since we do not have raw signals from the same sample for this dataset.

\begin{table}[tbh]
\centering
\caption{Accuracy and performance results when using R10.4 and R9.4 datasets}
\begin{tabular}{@{}llrr@{}}\toprule
\textbf{Flow Cell}   &         & \textbf{RH2} & \textbf{RH2-Min.}\\\midrule
\multicolumn{4}{c}{\textbf{Read Mapping Accuracy (E. coli)}} \\\midrule
                   & F1        & 0.9748          & 0.9631         \\
R9.4               & Precision & 0.9904          & 0.9865         \\
                   & Recall    & 0.9597          & 0.9408         \\\midrule
                   & F1        & 0.8960          & 0.8389         \\
R10.4              & Precision & 0.9506          & 0.9325         \\
                   & Recall    & 0.8473          & 0.7623         \\\midrule
\multicolumn{4}{c}{\textbf{Read Mapping Accuracy (S. aureus)}} \\\midrule
                   & F1        & 0.7749          & 0.6778         \\
R10.4              & Precision & 0.8649          & 0.8167         \\
                   & Recall    & 0.7018          & 0.5793         \\\midrule
\multicolumn{4}{c}{\textbf{Performance (E. coli)}} \\\midrule
R9.4               & Throughput [bp/sec]      & 303,382.45     & 659,013.57         \\
                   & Mean time per read [ms]  & 2.161          & 1.099         \\\midrule
R10.4              & Throughput [bp/sec]      & 175,351.94     & 480,471.75         \\
                   & Mean time per read [ms]  & 6.598          & 2.505         \\\midrule
\multicolumn{4}{c}{\textbf{Performance (S. aureus)}} \\\midrule
R10.4              & Throughput [bp/sec]      & 256,680.4      & 617,308.7         \\
                   & Mean time per read [ms]  & 5.478          & 2.243         \\\bottomrule
\end{tabular}

\label{supptab:r10accuracy}
\end{table}

\clearpage
\section{Performance} \label{suppsec:performance}
\subsection{Runtime, Peak Memory Usage, and Throughput} \label{suppsubsec:cpuandmemory}
Supplementary Table~\ref{supptab:performance} shows the computational resources required by each tool during the indexing and mapping steps. To measure the required computational resources, we collect CPU time and peak memory usage of each tool for all the datasets. To collect these results, we use \texttt{time -v} command in Linux. CPU time shows the total user and system time. Peak memory usage shows the maximum resident set size in the main memory that the application requires to complete its task. To measure the CPU threads needed for analyzing the entire MinION Flowcell with 512 pores, we divide 512 with the number of pores that a single thread can process (as shown with the values inside the bars in Figure~\ref{fig:throughput}) and round up the values to provide the maximum number of threads needed.

\begin{table}[tbh]
\centering
\caption{Computational resources required in the indexing step of each tool.}
\resizebox{0.55\columnwidth}{!}{
\begin{tabular}{@{}lrrrrr@{}}\toprule
\textbf{Dataset} & \textbf{RH2} & \textbf{RH2-Min.} & \textbf{RH} & \textbf{UNCALLED} & \textbf{Sigmap} \\\midrule
\multicolumn{6}{c}{Indexing CPU Time (sec)} \\\midrule
SARS-CoV-2 & 0.12 & 0.06 & 0.16 & 8.40 & \cellcolor{bestresult}\textbf{0.02} \\
E. coli & 2.48 & \cellcolor{bestresult}\textbf{1.61} & 2.56 & 10.57 & 8.86 \\
Yeast & 4.56 & \cellcolor{bestresult}\textbf{3.02} & 4.44 & 16.40 & 25.29 \\
Green Algae & 27.60 & \cellcolor{bestresult}\textbf{17.73} & 24.51 & 213.13 & 420.25 \\
Human & 1,093.56 & \cellcolor{bestresult}\textbf{588.30} & 809.08 & 3,496.76 & 41,993.26 \\
Contamination & 0.13 & 0.06 & 0.15 & 8.38 & \cellcolor{bestresult}\textbf{0.03} \\
Rel. Abundance & 747.74 & \cellcolor{bestresult}\textbf{468.14} & 751.67 & 3,666.14 & 36,216.87 \\
\midrule
\multicolumn{6}{c}{Indexing Peak Memory (GB)} \\\midrule
SARS-CoV-2 & \cellcolor{bestresult}\textbf{0.01} & \cellcolor{bestresult}\textbf{0.01} & \cellcolor{bestresult}\textbf{0.01} & 0.06 & \cellcolor{bestresult}\textbf{0.01} \\
E. coli & 0.35 & 0.19 & 0.35 & \cellcolor{bestresult}\textbf{0.11} & 0.40 \\
Yeast & 0.75 & 0.39 & 0.76 & \cellcolor{bestresult}\textbf{0.30} & 1.04 \\
Green Algae & 5.11 & \cellcolor{bestresult}\textbf{2.60} & 5.33 & 11.94 & 8.63 \\
Human & 80.75 & \cellcolor{bestresult}\textbf{40.59} & 83.09 & 48.43 & 227.77 \\
Contamination & \cellcolor{bestresult}\textbf{0.01} & \cellcolor{bestresult}\textbf{0.01} & \cellcolor{bestresult}\textbf{0.01} & 0.06 & \cellcolor{bestresult}\textbf{0.01} \\
Rel. Abundance & 152.59 & 75.62 & 152.84 & \cellcolor{bestresult}\textbf{47.80} & 238.32 \\
\midrule
\multicolumn{6}{c}{Mapping CPU Time (sec)} \\\midrule
SARS-CoV-2 & 1,705.43 & \cellcolor{bestresult}\textbf{1,227.05} & 1,539.64 & 29,282.90 & 1,413.32 \\
E. coli & 1,296.34 & \cellcolor{bestresult}\textbf{787.49} & 7,453.21 & 28,767.58 & 22,923.09 \\
Yeast & 545.77 & \cellcolor{bestresult}\textbf{246.37} & 4,145.38 & 7,181.44 & 7,146.32 \\
Green Algae & 2,135.83 & \cellcolor{bestresult}\textbf{657.63} & 22,103.03 & 12,593.01 & 26,778.44 \\
Human & 100,947.58 & \cellcolor{bestresult}\textbf{21,860.05} & 1,825,061.23 & 245,128.15 & 6,101,179.89 \\
Contamination & 3,783.69 & \cellcolor{bestresult}\textbf{2,332.28} & 3,480.43 & 234,199.60 & 3,011.78 \\
Rel. Abundance & 250,076.90 & \cellcolor{bestresult}\textbf{62,477.76} & 4,551,349.79 & 569,824.13 & 15,178,633.11 \\
\midrule
\multicolumn{6}{c}{Mapping Peak Memory (GB)} \\\midrule
SARS-CoV-2 & 4.15 & 4.16 & 4.20 & \cellcolor{bestresult}\textbf{0.17} & 28.26 \\
E. coli & 4.13 & 4.03 & 4.18 & \cellcolor{bestresult}\textbf{0.50} & 111.12 \\
Yeast & 4.38 & 4.12 & 4.37 & \cellcolor{bestresult}\textbf{0.36} & 14.66 \\
Green Algae & 6.11 & 4.98 & 11.77 & \cellcolor{bestresult}\textbf{0.78} & 29.18 \\
Human & 48.75 & 25.04 & 52.43 & \cellcolor{bestresult}\textbf{10.62} & 311.94 \\
Contamination & 4.16 & 4.14 & 4.17 & \cellcolor{bestresult}\textbf{0.62} & 111.70 \\
Rel. Abundance & 49.14 & 25.82 & 54.89 & \cellcolor{bestresult}\textbf{8.99} & 486.63 \\
\midrule
\multicolumn{6}{c}{Mapping Throughput (bp/sec)} \\\midrule
SARS-CoV-2 & 552,561.25 & \cellcolor{bestresult}\textbf{885,263.48} & 694,274.92 & 9,260.31 & 602,380.96 \\
E. coli & 303,382.45 & \cellcolor{bestresult}\textbf{659,013.57} & 72,281.32 & 7,515.76 & 13,750.97 \\
Yeast & 150,547.61 & \cellcolor{bestresult}\textbf{394,766.80} & 28,757.15 & 7,471.48 & 11,624.82 \\
Green Algae & 28,742.46 & \cellcolor{bestresult}\textbf{98,323.70} & 9,488.79 & 10,069.41 & 2,569.89 \\
Human & 8,968.78 & \cellcolor{bestresult}\textbf{37,086.38} & 2,099.35 & 7,225.67 & 236.45 \\
Contamination & 563,129.81 & \cellcolor{bestresult}\textbf{884,929.30} & 696,873.20 & 9,343.95 & 601,936.49 \\
Rel. Abundance & 9,501.37 & \cellcolor{bestresult}\textbf{36,919.79} & 962.79 & 8,437.70 & 196.48 \\
\midrule
\multicolumn{6}{c}{CPU Threads Needed for the entire MinION Flowcell (512 pores)} \\\midrule
SARS-CoV-2 & \cellcolor{bestresult}\textbf{1} & \cellcolor{bestresult}\textbf{1} & \cellcolor{bestresult}\textbf{1} & 25 & \cellcolor{bestresult}\textbf{1} \\
E. coli & \cellcolor{bestresult}\textbf{1} & \cellcolor{bestresult}\textbf{1} & 4 & 31 & 17 \\
Yeast & 2 & \cellcolor{bestresult}\textbf{1} & 9 & 31 & 20 \\
Green Algae & 9 & \cellcolor{bestresult}\textbf{3} & 25 & 23 & 90 \\
Human & 26 & \cellcolor{bestresult}\textbf{7} & 110 & 32 & 975 \\
Contamination & \cellcolor{bestresult}\textbf{1} & \cellcolor{bestresult}\textbf{1} & \cellcolor{bestresult}\textbf{1} & 25 & \cellcolor{bestresult}\textbf{1} \\
Rel. Abundance & 25 & \cellcolor{bestresult}\textbf{7} & 240 & 28 & 1173 \\\bottomrule
\multicolumn{6}{l}{\footnotesize Best results are \colorbox{bestresult}{\textbf{highlighted}}.} \\
\end{tabular}

}
\label{supptab:performance}
\end{table}

\clearpage
\subsection{Impact of Different File Formats on Performance} \label{suppsubsec:fileformats}

Supplementary Table~\ref{supptab:pod5_fast5_resources} shows the overall execution time when using different raw signal file formats: FAST5, POD5, and BLOW5~\citesupp{supp_gamaarachchi_fast_2022}. To evaluate the direct impact of these formats, we run \mech (RH2) and \mech-Minimizer (RH2-Min.) 1)~using a single thread (i.e., single thread for the entire execution including \emph{both} file IO and mapping), 2)~using an isolated SSD on a PCI-e interface, 3)~using the same compression type (i.e., zstd) for all file formats, and 4)~clearing the disk cache before each execution. When using a single thread, we confirm that the underlying libraries for FAST5, POD5, and BLOW5 are not aggressively using more threads than what is allocated to them, as the thread utilization is reported as 0.99 (i.e., 99\%) by the \texttt{time -v} command for the entire execution.

We note that even if we use multiple threads when running \mech, the file IO step (i.e., reading from or writing to a file) always uses a single thread and is overlapped with the mapping step (i.e., either the read or write operation is run in parallel together with the mapping step by using one thread where the mapping step takes rest of the allocated threads). The design is due to the pipelining implementation strategy we adopt, similar to the minimap2 implementation~\cite{li_minimap2_2018}. We note that if \mech is run using a single thread, none of these steps overlap with each other, and they run sequentially using only one thread, which is our evaluation setting we show in Supplementary Table~\ref{supptab:pod5_fast5_resources}.

\begin{table}[tbh]
\centering
\caption{Comparison of overall execution time when using different file formats in \mech in a single-threaded mode.}
\begin{tabular}{@{}lrr@{}}\toprule
\textbf{Tool} & \textbf{\emph{E. coli}} & \textbf{\emph{Yeast}}  \\\midrule
\multicolumn{3}{c}{Elapsed Time (mm:ss)} \\\midrule
RH2-FAST5   & 19:27 & 08:35  \\
RH2-POD5    & 16:55 & 07:33  \\
RH2-BLOW5    & 17:32 & 07:38  \\\midrule
RH2-Min.-FAST5   & 12:13 & 03:56  \\
RH2-Min.-POD5    & 09:42 & 02:56  \\
RH2-Min.-BLOW5    & 10:16 & 03:02  \\\midrule
\end{tabular}

\label{supptab:pod5_fast5_resources}
\end{table}

\subsection{Mapping Time per Read}\label{subsec:mappingtimeperread}

\begin{figure}[tbh]
  \centering
  \includegraphics[width=0.8\columnwidth]{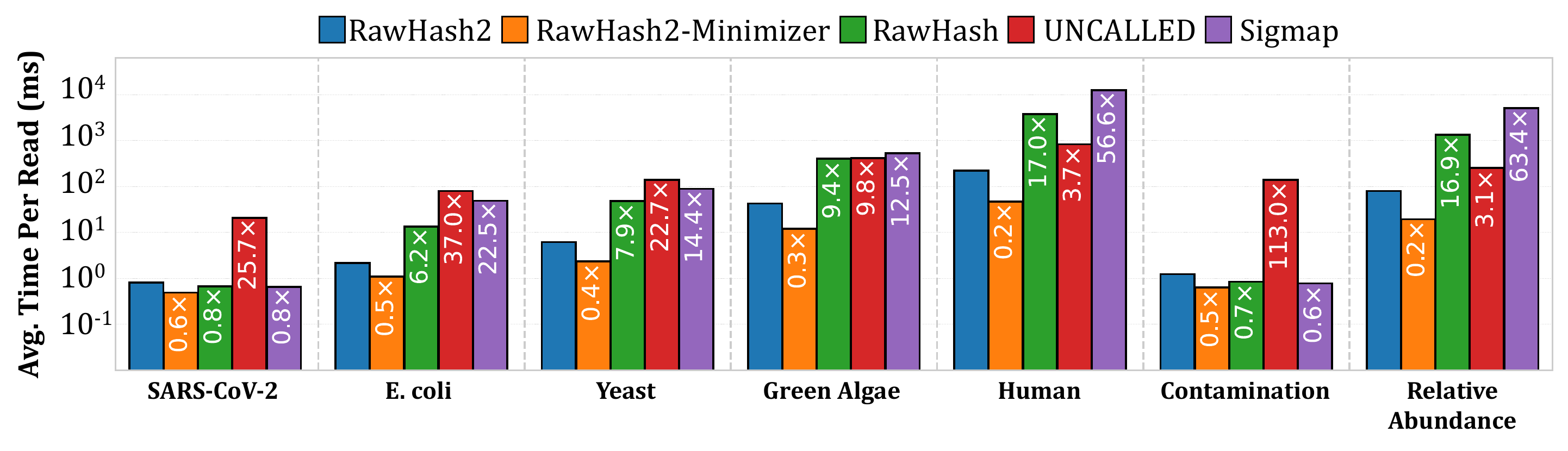}
  \caption{Average time spent per read by each tool in real-time. Values inside the bars show the speedups that \mech provides over other tools in each dataset.}
  \label{suppfig:timeperread}
\end{figure}

Supplementary Figure~\ref{suppfig:timeperread} shows the average mapping time that each tool spends per read for all the datasets we evaluate. The mapping times spent per read are provided by each tool as PAF output with the \texttt{mt} tag. We use these reported values to calculate the average mapping time across all reads reported in their corresponding PAF files.

\clearpage

\subsection{Combined benefits of performance, accuracy, and average sequencing length}\label{suppsubsec:combined}

\begin{figure}[tbh]
\centering
\includegraphics[width=\columnwidth]{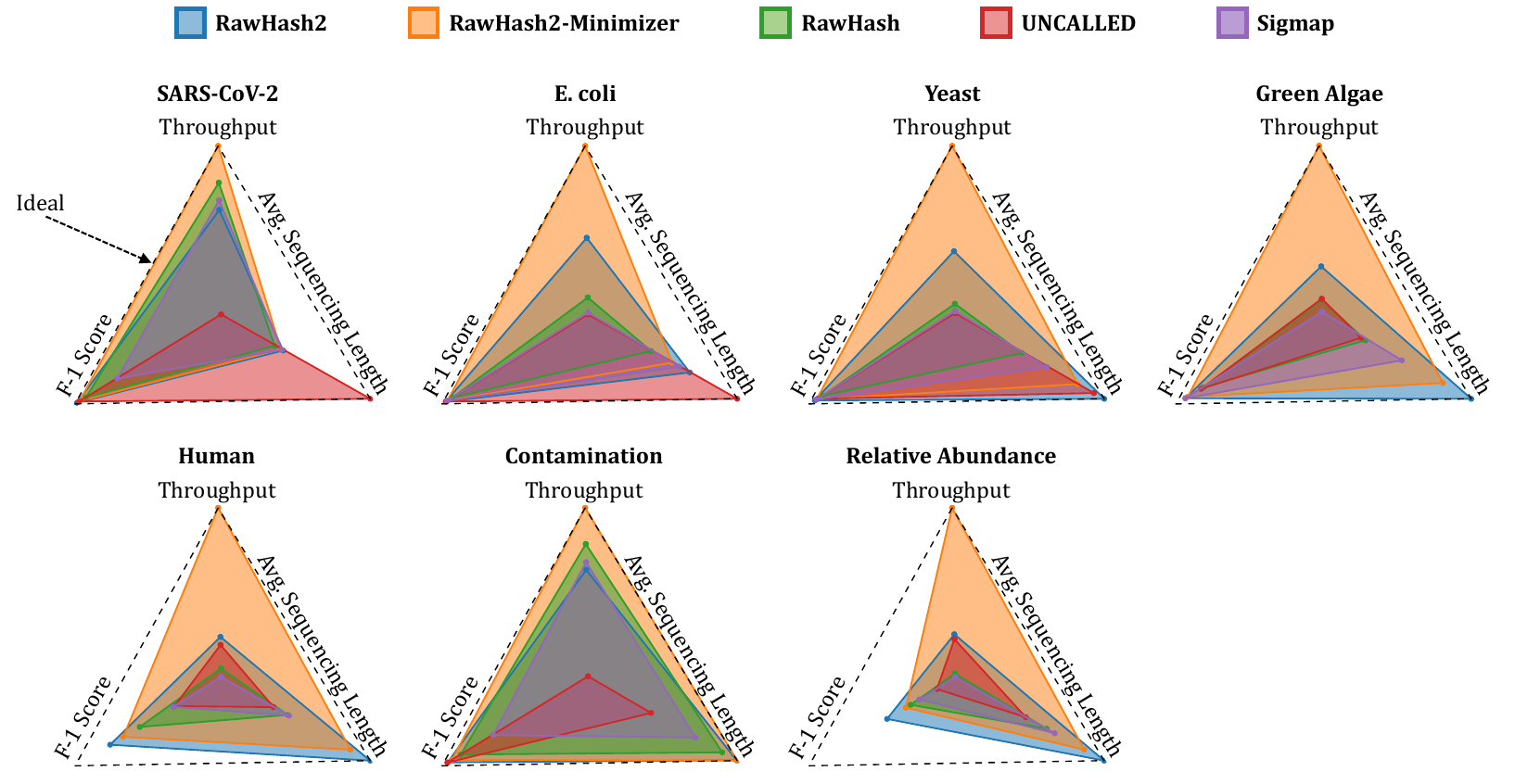}
\caption{Combined results in terms of throughput, F-1 score (i.e., accuracy), and average sequencing length across different datasets. The dotted triangles show the best possible results, where each edge shows the best result for its corresponding metric.}
\label{suppfig:combined}
\end{figure}

Supplementary Figure~\ref{suppfig:combined} shows the combined results of each tool in terms of throughput, F-1 Score (i.e., accuracy), and average sequencing length for each dataset. The dotted lines in each triangle show the ideal combined result. Each edge of the triangle shows the best result for the corresponding metric, as shown in the figure.

For the edge that shows the F-1 score, the best point is 1.0. All tools have F-1 scores between 0 and 1, as shown in Table~\ref{tab:accuracy}. For the other two edges, which show throughput and average sequencing length, the best result is determined based on the highest result we observe for that dataset. We adjust all other results using these highest results so that the adjusted throughput and average sequencing length values are always between 0 and 1.

\subsection{Ratio of Filtered Seeds from Frequency Filter}\label{suppsubsec:freqfilter}

Supplementary Table~\ref{supptab:filtered_seed_hits} shows the ratio of seed hits filtered out by the frequency filtered in \mech. We calculate these ratios in three steps.
First, for each seed (i.e., a hash value that \mech constructs from raw signals), we perform a query to the hash table that is used as an index. If the hash value exists, the table returns a list of genomic regions that share the same hash value. Each region counts as a seed hit, and the list length indicates the number of seed hits.
Second, for all seeds generated from raw signals, we count 1)~the overall number of seed hits and 2)~the number of seed hits filtered out by frequency filter. We note that if the list length (i.e., number of seed hits) returned after querying a particular seed is above a certain threshold (defined by our frequency filter), all seed hits within the same list are filtered out.
Third, we calculate the ratio of filtered seed hits to the total seed hits and report these ratios in Supplementary Table~\ref{supptab:filtered_seed_hits}.

\begin{table}[tbh]
\centering
\caption{Ratio of filtered seed hits from frequency filter.}
\begin{tabular}{@{}lr@{}}
\toprule
\textbf{Dataset}      & \textbf{Average Filtered Ratio} \\ \midrule
SARS-CoV-2            & 0.0627                       \\
E. coli        & 0.5505                       \\
Yeast                 & 0.5356                       \\
Green Algae           & 0.8106                       \\
Human & 0.5104                       \\
E. coli (R10.4)       & 0.6895                      \\
S. aureus (R10.4)     & 0.6003                       \\ \bottomrule
\end{tabular}

\label{supptab:filtered_seed_hits}
\end{table}

\clearpage
\section{Configuration} \label{sec:configuration}
\subsection{Datasets} \label{subsec:datasets}

In Supplementary Table~\ref{supptab:dataset} we show the details of the datasets used in our evaluation and their corresponding sequencing run settings. The \emph{Basecaller Model} column shows the details about the basecaller model and the version we use. Except for the D7 dataset, all other datasets include the basecalled sequences within their corresponding FAST5 files or the corresponding accession numbers available at NCBI. We provide the scripts to extract these basecalled sequences on the GitHub page of \mech. For the D7 dataset, we provide the necessary commands to run dorado for basecalling on the GitHub page.

\begin{table}[tbh]
\centering
\caption{Details of datasets used in our evaluation.}
\resizebox{\columnwidth}{!}{\begin{tabular}{@{}cllllllrrllr@{}}\toprule
& \textbf{Organism}     & \textbf{Device} & \textbf{Flow Cell} & \textbf{Transloc.} & \textbf{Sampling} & \textbf{Basecaller}     &\textbf{Reads} & \textbf{Bases}              & \textbf{SRA}       & \textbf{Reference}      & \textbf{Genome}\\
& \textbf{}             & \textbf{Type} & \textbf{Type}     & \textbf{Speed}  & \textbf{Frequency} &        \textbf{Model}       &\textbf{(\#)}  & \textbf{(\#)}           & \textbf{Accession} & \textbf{Genome}         & \textbf{Size}  \\\midrule
\multicolumn{12}{c}{Read Mapping} \\\midrule
D1 & \emph{SARS-CoV-2}  & MinION & R9.4.1 e8 (FLO-MIN106) & 450 & 4000 & Guppy HAC v3.2.6 & 1,382,016      & 594M               & CADDE Centre & GCF\_009858895.2   & 29,903 \\\midrule
D2 & \emph{E. coli}     & GridION & R9.4.1 e8 (FLO-MIN106) & 450 & 4000 & Guppy HAC v5.0.12  & 353,317        & 2,365M                 & ERR9127551         & GCA\_000007445.1   & 5M \\\midrule
D3 & \emph{Yeast}       & MinION & R9.4.1 e8 (FLO-MIN106) & 450 & 4000 & Albacore v2.1.7 & 49,989         & 380M                   & SRR8648503         & GCA\_000146045.2   & 12M\\\midrule
D4 & \emph{Green Algae} & PromethION & R9.4.1 e8 (FLO-PRO002) & 450 & 4000 & Albacore v2.3.1 & 29,933         & 609M                   & ERR3237140         & GCF\_000002595.2   & 111M\\\midrule
D5 & \emph{Human} & MinION & R9.4.1 e8 (FLO-MIN106) & 450 & 4000 & Guppy Flip-Flop v2.3.8 & 269,507        & 1,584M                 & FAB42260 & T2T-CHM13 (v2)     & 3,117M\\\midrule
D6 & \emph{E. coli}     & GridION & R10.4 e8.1 (FLO-MIN112) & 450 & 4000 & Guppy HAC v5.0.16 & 1,172,775        & 6,123M       & ERR9127552         & GCA\_000007445.1   & 5M \\\midrule
D7 & \emph{S. aureus}   & GridION & R10.4 e8.1 (FLO-MIN112) & 450 & 4000 & Dorado SUP v0.5.3 & 407,727        & 1,281M       & SRR21386013         & GCF\_000144955.2   & 2.8M \\\midrule
\multicolumn{12}{c}{Contamination Analysis} \\\midrule
\multicolumn{7}{c|}{D1 and D5} & 1,651,523    & 2,178M                 & D1 and D5             & D1                 & 29,903\\\midrule
\multicolumn{12}{c}{Relative Abundance Estimation} \\\midrule
\multicolumn{7}{c|}{D1-D5} & 2,084,762    & 5,531M                  & D1-D5              & D1-D5              & 3,246M\\\bottomrule
\multicolumn{12}{l}{Multiple dataset numbers in contamination analysis and relative abundance estimation show the combined datasets.}\\
\multicolumn{12}{l}{D1-D5 datasets are from R9.4, and D6 and D7 are from R10.4. Human reads are from Nanopore WGS.}\\
\multicolumn{12}{l}{Base counts in millions (M).}\\
\end{tabular}
}
\label{supptab:dataset}
\end{table}

\subsection{Parameters} \label{subsec:parameters}

In Supplementary Table~\ref{tab:parameters}, we show the parameters of each tool for each dataset. In Supplementary Table~\ref{tab:presets}, we show the details of the preset values that \mech sets in Supplementary Table~\ref{tab:parameters}. For \unc~\citesupp{supp_kovaka_targeted_2021}, \sig~\citesupp{supp_zhang_real-time_2021}, and minimap2~\citesupp{supp_li_minimap2_2018}, we use the same parameter setting for all datasets. For the sake of simplicity, we only show the parameters we explicitly set in each tool. For the descriptions of all the other parameters, we refer to the help message that each tool generates, including \mech.

\begin{table}[tbh]
\centering
\caption{Parameters we use in our evaluation for each tool and dataset in mapping.}
\resizebox{\linewidth}{!}{
\begin{tabular}{@{}lccccccccc@{}}\toprule
\textbf{Tool} & \textbf{\emph{Contamination}} & \textbf{\emph{SARS-CoV-2}} & \textbf{\emph{E. coli} (R9.4)} & \textbf{\emph{Yeast}} & \textbf{\emph{Green Algae}} & \textbf{\emph{Human}} & \textbf{\emph{Rel. Abundance}} & \textbf{\emph{E. coli} (R10.4)} & \textbf{\emph{S. aureus} (R10.4)} \\\midrule
\mech    & -x viral --depletion -t 32 & -x viral -t 32 & -x sensitive -t 32 & -x sensitive -t 32 & -x sensitive -t 32 & -x fast -t 32 & -x fast -t 32 & -x sensitive --r10 -t 32 & -x sensitive --r10 -t 32\\\midrule
\rhmin   & -x viral -w3 --depletion -t 32  & -x viral -w3 -t 32  & -x sensitive -w3 -t 32 & -x sensitive -w3 -t 32 & -x sensitive -w3 -t 32 & -x fast -w3 -t 32 & -x fast -w3 -t 32 & -x sensitive --r10 -w3 -t 32 & -x sensitive --r10 -w3 -t 32\\\midrule
\rh         	 & -x viral -t 32  & -x viral -t 32  & -x sensitive -t 32 & -x sensitive -t 32 & -x fast -t 32 & -x fast -t 32 & -x fast -t 32 & NA & NA\\\midrule
\unc  			 & \multicolumn{7}{c}{map -t 32} & NA & NA\\\midrule
\sig        	 & \multicolumn{7}{c}{-m -t 32} & NA & NA\\\midrule
Minimap2         & \multicolumn{9}{c}{-x map-ont -t 32}\\\bottomrule
\end{tabular}

}
\label{tab:parameters}
\end{table}

\begin{table}[tbh]
\centering
\caption{Corresponding parameters of presets (-x) in \mech.}
\resizebox{\linewidth}{!}{
\begin{tabular}{@{}lcc@{}}\toprule
\textbf{Preset} & \textbf{Corresponding parameters} & Usage \\\midrule
viral      & -e 6 -q 4 --max-chunks 5 --bw 100 --max-target-gap 500 & Viral genomes\\
& --max-target-gap 500 --min-score 10 --chain-gap-scale 1.2 --chain-skip-scale 0.3 &  \\\midrule
sensitive  & -e 8 -q 4 --fine-range 0.4 & Small genomes (i.e., $<500M$ bases)\\\midrule
fast       & -e 8 -q 4 --max-chunks 20 & Large genomes (i.e., $>500M$ bases)\\\midrule
\multicolumn{3}{c}{\textbf{Other helper parameters}}\\\midrule
depletion  & --best-chains 5 --min-mapq 10 --w-threshold 0.5 & Contamination analysis\\
& --min-anchors 2 --min-score 15 --chain-skip-scale 0 & \\\midrule
r10        & -k9 --seg-window-length1 3 --seg-window-length2 6 --seg-threshold1 6.5 & For R10.4 Flow Cells \\
& --seg-threshold2 4 --seg-peak-height 0.2 --chain-gap-scale 1.2 & \\\bottomrule
\end{tabular}

}
\label{tab:presets}
\end{table}

\clearpage

\subsection{Versions}\label{subsec:versions}

Supplementary Table~\ref{tab:versions} shows the version and the link to these corresponding versions of each tool and library we use in our experiments and in \mech, respectively.

\begin{table}[tbh]
\centering
\caption{Versions of each tool and library.}
\begin{tabular}{@{}lll@{}}\toprule
\textbf{Tool} & \textbf{Version} & \textbf{Link to the Source Code} \\\midrule
\mech & 2.1 & \url{https://github.com/CMU-SAFARI/RawHash/releases/tag/v2.1}\\\midrule
\rh & 1.0 & \url{https://github.com/CMU-SAFARI/RawHash/releases/tag/v1.0}\\\midrule
\unc  & 2.3 & \url{https://github.com/skovaka/UNCALLED/releases/tag/v2.3}\\\midrule
\sig  & 0.1 & \url{https://github.com/haowenz/sigmap/releases/tag/v0.1}\\\midrule
Minimap2 & 2.24 & \url{https://github.com/lh3/minimap2/releases/tag/v2.24}\\\midrule
\multicolumn{3}{c}{Library versions}\\\midrule
FAST5 (HDF5) & 1.10 & \url{https://github.com/HDFGroup/hdf5/tree/db30c2d} \\\midrule
POD5 & 0.2.2 & \url{https://github.com/nanoporetech/pod5-file-format/releases/tag/0.3.10}\\\midrule
S/BLOW5 & 1.2.0-beta & \url{https://github.com/hasindu2008/slow5lib/tree/e0d0d0f}\\\bottomrule
\end{tabular}

\label{tab:versions}
\end{table}

\clearpage

\let\noopsort\undefined
\let\printfirst\undefined
\let\singleletter\undefined
\let\switchargs\undefined

\bibliographystylesupp{IEEEtran}
\bibliographysupp{main}

\end{document}